\documentclass[10pt,A4,twocolumn]{article}
\usepackage[T1]{fontenc}
\usepackage{times,graphicx,units,amsfonts}
\usepackage{natbib}
\begin{document}
\title{Penetrative turbulence associated with mesoscale surface heat flux variations}
\author{Jahrul M Alam\thanks{Corresponding author's email: alamj@mun.ca} and M Alamgir Hossain\\
Department of Mathematics and Statistics\\
Memorial University, Canada}
\date{}
    \maketitle
    \begin{abstract}
      This article investigates penetrative turbulence in the atmospheric boundary layer. Using a large eddy simulation approach, we study characteristics of the mixed layer with respect to surface heat flux variations in the range from $231.48~\unit{W m}^{-2}$ to $925.92~\unit{W m}^{-2}$, and observe that the surface heterogeneity on a spatial scale of $20$~km leads to downscale turbulent kinetic energy cascade. Coherent fluctuations of mesoscale horizontal wind is observed at $100$~m above the ground. Such a surface induced temporal oscillations in the horizontal wind suggest a rapid jump in mesocale wind forecasts, which is difficult to parameterize using traditional one-dimensional ensemble-mean models. Although the present work is idealized at a typical scale ($20$~km) of surface heterogeneity, the results help develop effective subgrid scale parameterization schemes for classical weather forecasting mesoscale models.
    \end{abstract}

\section{Introduction}
Earth's surface is assumed to play a distinctive influence in weather and climate systems~\cite[][]{Stull76}. The surface influence generates horizontal convection as a result of differential heating along one horizontal boundary of the atmospheric boundary layer (ABL)~\cite[][]{Lane2008,Scotti2011}. Coherent mesoscale motions are reported by many studies, where the effect of surface heterogeneity are investigated numerically on scales of few hundred kilometers, using a grid spacing of $\Delta x\sim\mathcal O(10~\unit{km})$~\cite[][]{Skamarock2008,Lin2007}. However, in the geophysical fluid dynamics community, controversial opinions exist on whether horizontal convection due to differential heating at the same level would lead to turbulence, or drive large scale overturning circulations~\cite[][]{Paparella2002,Scotti2011}. A century old experimental result by~\cite{Sand1908} leads to the opinion that a sustained circulation cannot occur if the level of the heating source is the same as that of the cooling source. This hypothesis is supported theoretically by the anti-turbulence theorem of~\cite{Paparella2002}. However, \cite{Scotti2011} employed a direct numerical simulation of horizontal convection at Rayleigh number $10^{10}$ to show that a flow driven by the horizontal convection exhibits the characteristics of a true turbulent flow. Clearly, some fundamental aspects of solar heating at the earth's surface is not fully understood. 

Much of our current understanding of surface induced phenomena depends on numerical simulations and measurements/observations. For example, lightning data over Houston, USA between years of 1989 and 2000 indicates that highest flash densities occur over urban areas. Measurements of turbulence for the $43$~years period from $1958$ to $2001$ in the North Atlantic, USA, and European sectors conclude that clear-air turbulence increases in these regions by $40-90$\% as a result of the urban induced impact on vertical transport and mixing of penetrative ABL turbulence. Note that the average horizontal flux of kinetic energy at a height~$\sim 100$~m from the ground is about $1\,000$~W~m$^{-2}$. A destabilization of this huge energy flux due to differential heating on the surface may cause catastrophic impact on the turbulent atmosphere.  
Since penetrative turbulence in the atmosphere~\cite[see][]{Stull76} is primarily characterized by solar heating, perturbations to boundary layer structures by human activities ({\em e.g.} urbanization) are sensitive to vertical mixing and transport~\cite[][]{Bryan2002,Lane2008}. 

This article reports on a large eddy simulation (LES) based numerical model in order to characterize penetrative turbulence over heterogeneous land surface. Primary objectives of this short article includes the investigation of evidences whether horizontal convection transports large quantities of heat, as well as sustains large amounts of diapycnal mixing with a relatively small amount of dissipation. We demonstrate that differential surface heating on scales of a typical modern city cascades kinetic energy downscale, which -- in the absence of shear -- is sufficient to initiate a turbulent flow. 
The computational fluid dynamics approach used in this work is fully detailed by~\citet{Alam2014}. This method filters the governing equations using a multiresolution approach, and employs a Smagorinsky type eddy viscosity model for the subfilter scale processes. We use a sixth order weighted residual collocation method that has no inherent numerical dissipation, and thus, there is no need to apply artificial dissipation to ensure numerical stability.

\section{Methodology}
\subsection{Theory}
Penetrative turbulence occurs when an unstably stratified large body of fluid underlies a stably stratified fluid layer. In this situation, turbulent eddies are driven vertically by the buoyancy force, and attempt to penetrate into the overlying stable fluid~\cite[][]{Stull76}. In the stable region, when an eddy reaches its level of buoyancy, it returns back downward into the unstable region. However, due to the gained momentum, eddies often overshoot their level of buoyancy, and thus, internal waves are excited from the interface between the overlying stably stratified fluid and the underling unstable mixing layer~\cite[][]{Stull76,Lane2008}. These waves transports kinetic energy to the upper atmosphere, and may initiate upper level clear air turbulence. \citet{Deardorff69} investigated such penetrative turbulence experimentally. Here, we develop a large eddy simulation model to simulate the above mentioned phenomena.

\subsection{Equations}
For a compressible atmospheric model, the temperature ($T$) and the pressure ($p$) are represented by the potential temperature $(\theta)$ and the Exner function $(\pi')$ \cite[e.g.,][]{Piotr2014},
$$
\theta = T\left(\frac{p_0}{p}\right)^{R_d/c_p},
\quad
\pi' = \left(\frac{R_d}{p_0}\rho\theta\right)^{R_d/(c_p-R_d)},
$$
respectively, where $p_0$ is a reference pressure, $R_d$ is the gas constant, $c_p$ is the specific heat at constant pressure, and $\rho$ is the density. The continuity equation is replaced with Eq~(\ref{eq:pi}). The notation $(x_1,x_3)=(x,z)$ and $(u_1,u_3)=(u,w)$ are adopted for simplicity. A spatial filter is applied to the momentum and energy equations, where $u_i(\equiv\langle u_i\rangle)$ and $\theta(\equiv\langle\theta\rangle)$ represent filtered velocity and temperature, respectively. More specifically, $\tilde u_i = u_i +u'_i$ and $\tilde\theta = \theta_0 + \bar\theta(z) + \theta + \theta'$~\cite[][]{dear70,Deardorff80}. Note the separation of the reference temperature $\theta_0$ from the background temperature $\bar\theta(z)$, which is convenient for satisfying the surface condition. 
The following equations are solved in the present work;
\begin{equation}
  \label{eq:pi}
  \frac{D\pi'}{Dt} = -\pi'\frac{\partial u_i}{\partial x_i}, 
\end{equation}
\begin{equation}
  \label{eq:me}
  \frac{Du_i}{Dt} = -c_p\theta_0\frac{\partial\pi'}{\partial x_i}+\frac{g\theta}{\theta_0}\delta_{i3}  -\frac{\partial\tau_{ij}}{\partial x_j},
\end{equation}
\begin{equation}
  \label{eq:vpr}
  \frac{D\theta}{Dt} = -w\frac{\partial\bar\theta}{\partial z}-\frac{\partial\tau_{\theta j}}{\partial x_j}.
\end{equation}
The simulation region is a vertical plane $(x,z)$ that extends $100$~km horizontally ($-50\le x\le 50$) and $2$~km vertically ($0\le z\le 2$). A city of scale $20$~km exists for $-10\le x \le 10$, which is surrounded by rural areas. Recent literature  indicates that the most appropriate model for the near surface penetrative turbulence is not fully understood. As a compromise, as discussed by~\cite{Pope2000}, we have adopted a resolution that is finer than that is used by LES models of the ABL. The finest resolution uses $\Delta x=97.65$~m and $\Delta z=3.9$~m. Thus, a significant fraction of the energy containing large eddies is resolved. Here, some advantages of the costly three-dimensional simulation are sacrificed for a high resolution two-dimensional idealization~\cite[][]{Lane2008}.

%
\subsection{Large eddy simulation}
The subgrid scale turbulent stress is estimated by the popular Smagorinsky (1963) model, $\tau_{ij}=-2(C_s\Delta)^2|S|S_{ij}$, where $|S|=\sqrt{2S_{ij}S_{ij}}$ and $S_{ij}=(1/2)(\frac{\partial u_i}{\partial x_j} + \frac{\partial u_i}{\partial x_j})$. We define the filter width by $\Delta = \sqrt{\Delta x\cdot\Delta z}$, and take $C_s=0.18$ for the Smagorinsky constant. 

According to~\cite{Deardorff80}, the subgrid scale eddy coefficients may be computed by $K_h=(1+2l/\Delta)K_m$ and $K_m = 2(C_s\Delta)^2|S|$, where  $l\le\Delta$ is a subgrid scale mixing length. The length scale, $l$, is related to the subgrid scale turbulence energy $e'$ and the buoyancy frequency, $N^2=\frac{g}{\theta_0}\frac{\partial\bar\theta}{\partial z}$, {\em i.e.} $l=0.76\sqrt{e'}/N$. If the scale of resolved eddies is $\Delta=20$~m, then a turbulent Prandtle number $\mathcal Pr=0.71$ gives $l\approx 4$~m. The SGS flux for buoyancy is $\tau_{\theta i} = -K_h\frac{\partial\theta}{\partial x_i}$~\cite[][]{Deardorff80}.

\subsection{Surface heat-flux variation}\label{sec:sfc}
We consider a time independent profile for the surface heat flux variation $H_{\hbox{sfc}}(x) = \langle H_{\hbox{sfc}}\rangle +H_0$ that is given by
$$
H_{\hbox{sfc}}(x) = \langle H_{\hbox{sfc}}\rangle + A[\tanh\xi(x+\lambda/2) - \tanh\xi(x-\lambda/2)].
$$
Here, $\langle H_{\hbox{sfc}}\rangle=0.03~\hbox{K m s}^{-1}$ (about $37\hbox{ W m}^{-2}$) represents a domain average heat flux for all $x$ in the range from $-50$ to $50$~km, where there is no city induced heat flux $H_0$. $\lambda$ is the characteristic wavelength for the surface heat flux variation $H_0$, where the heating region (city) of the surface is from $-10$ to $10$~km, {\em i.e.} $\lambda=20$~km. This choice for $\lambda$ represents the resolved scale for mesoscale numerical weather prediction models. $\xi=100$ is a dimensionless number that leads to a continuous sharp interface between the central heating region and other part of the surface. Thus, the surface heat flux $H_{\hbox{sfc}}(x)$ takes approximately the form of a square wave without sharp corners, and models the effect of urban-rural heat flux variation. This article summarizes the sensitivity of surface heat flux on penetrative turbulence for $6$ values of $H_0$.  A purpose of these simulations is to understand the sensitivity of high-amplitude surface heat-flux heterogeneity for temporal oscillation in mesoscale atmospheric circulations. 

\subsection{Validation}
The total heat flux is composed of turbulent ($\overline{w'\theta'}$) and viscous ($\alpha\frac{\partial\theta}{\partial z}$) components, where the turbulent component vanishes on a flat surface. The dimensionless surface heat flux takes the form $-\frac{1}{\sqrt{RaPr}}\frac{\partial\theta}{\partial z}|_{z=0}$, where the potential temperature is the same the surface temperature.  

A scale analysis is necessary to compare the present model with that presented by~\cite{Dubois2009}. With a fully developed turbulence in the convective boundary layer, the vertical temperature profile satisfies $\frac{\partial\theta}{\partial z}=0$ in mixing region ($z>0$)~\cite[][]{Deardorff69}. In this case, natural convection heat transfer is characterized by the Rayleigh number $\mathcal Ra = \frac{gH^3}{\theta_0\nu\kappa}\frac{H_0H}{\alpha}$, where $H$ is a vertical length scale, $1/\theta_0$ is the coefficient of thermal expansion (1/K), $\nu$ is the kinematic viscosity~(m$^2$/s), $\kappa$ is thermal conductivity (W/m$\cdot$K), $g$ is the acceleration due to gravity~(m/s$^2$), and $\alpha$ is the thermal diffusivity~(m$^2$/s). In an LES, $\nu$ and $\alpha$ can be replaced with $K_m$ and $K_h$, respectively. So, we set $\frac{H_0H}{\alpha}=10^{\hbox{\tiny o}}$K to match the adiabatic lapse rate ($10^{\hbox{\tiny o}}$K/km) of the atmosphere so that $\frac{\partial\theta}{\partial z}=0$. As a result, an increase of $H_0$ by a factor of $2$ increases $\mathcal Ra$ by a factor of $10$. Thus, the values of $H_0=57.87~\unit{Wm^{-2}}$ and $115.74~\unit{Wm^{-2}}$ represents $\mathcal Ra=10^4$ and $10^5$, respectively. The comparison is summarized in Table~\ref{tab:cdt}. In addition, we have compared the vertical profile of mean temperature with that obtained from the Wangara day 33 experiment, which shows an excellent agreement on  $\frac{\partial\theta}{\partial z}=0$ between two data sets. 

\begin{table}[b]
\centering
\begin{tabular}{|c|r|r|r|r|r|r|}
\hline
$H_0$ & \multicolumn{2}{ |c| }{ $57.87~\unit{Wm^{-2}}$} &\multicolumn{2}{ |c| }{ $115.74~\unit{Wm^{-2}}$}\\
\hline \hline
&Present&D \& T&Present&D \& T\\
\hline
$\theta_{\min}$ &  -0.0644 &-0.0712 &  -0.1672  & -0.1663\\
\hline
$u_{\max}$ &  0.1761 &0.1748 & 0.1796 &0.1790\\
\hline
$w_{\max}$ & 0.2275 & 0.2282&   0.3294 & 0.3224\\
\hline
$Nu$ & 0.3262 & 0.2951 & 0.6892 & 0.6435 \\

\hline
\end{tabular}
\caption {Comparison of dimensionless extreme values with that from \cite{Dubois2009} (D \& T).}
\label{tab:cdt}
\end{table}

\section{Results}
\subsection{Surface induced impact of air pollution}
\begin{figure*}[ht]
  \centering
  \begin{tabular}{cc}
  \includegraphics[height=2.7cm]{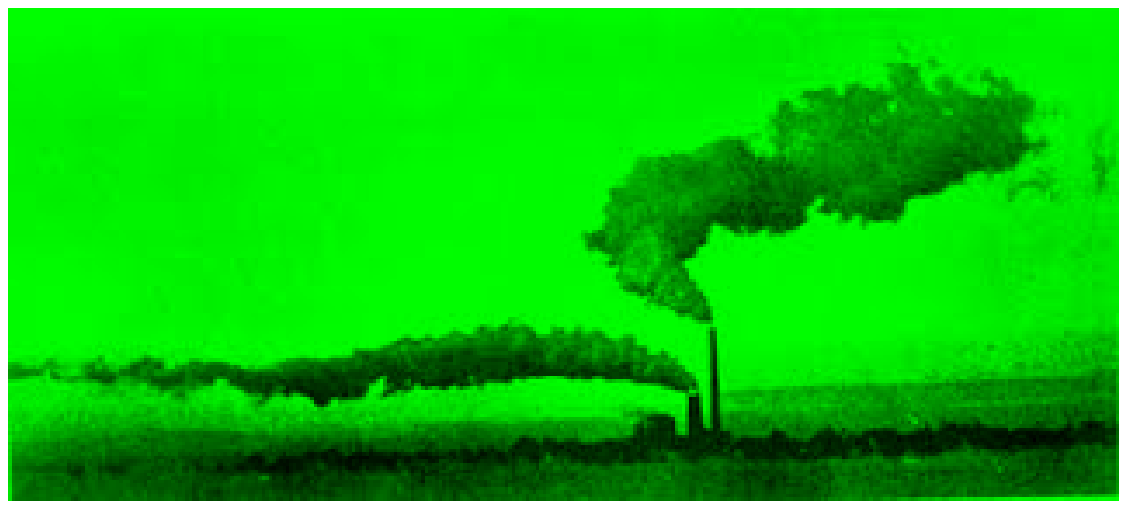}&
  \includegraphics[height=3cm]{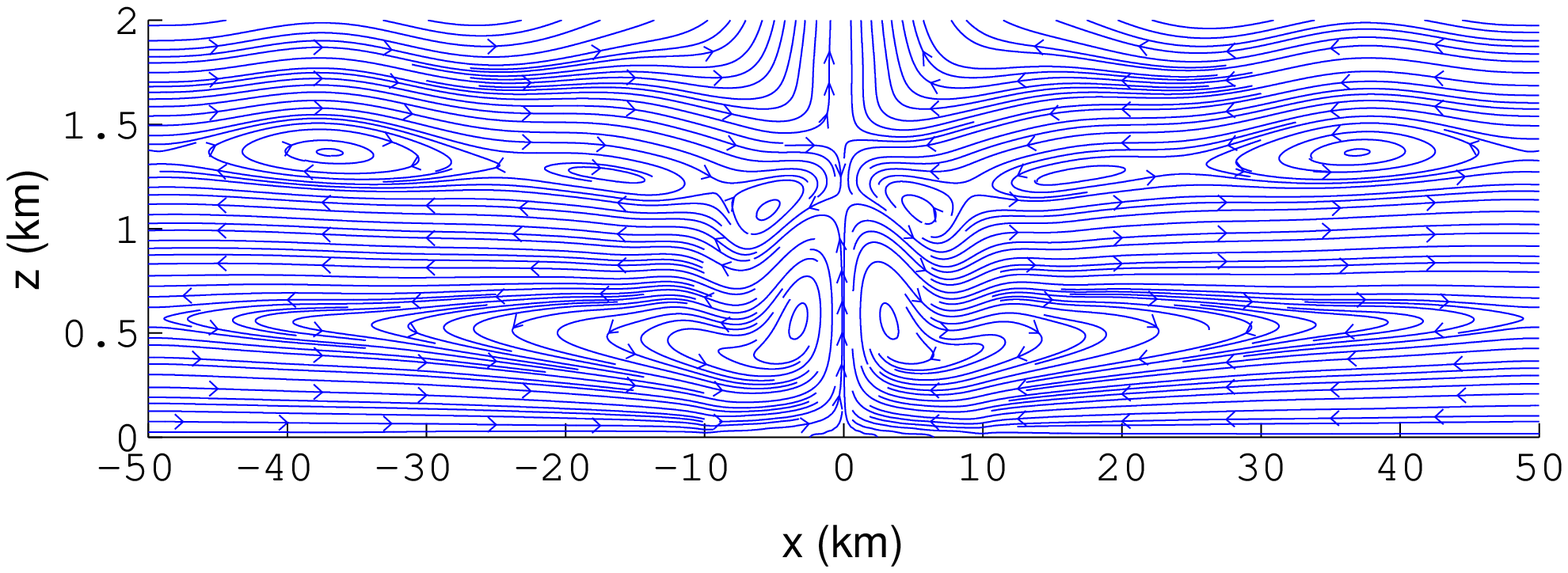}\\
  $(a)$&
  $(b)$
  \end{tabular}
  \caption{$(a)$~Movement of plumes from two chimneys of unequal height as a result of differential surface heating. The figure is adapted from Google image database, and represents a pattern of expected circulation. $(b)$~Low level converging flow and high level diverging flow is seen from this streamline at $t=6.5$~h for $H_0=115.74\hbox{ W m}^{-2}$.}
  \label{fig:sb}
\end{figure*}

Fig~\ref{fig:sb}$(a)$ demonstrates an example of surface induced impact on air pollution. The turbulent plume from the lower chimney ($75$~m tall) moves toward the region of warmer surface (city area), and that from the higher  chimney ($150$~m tall) moves toward the region of cooler surface (urban, ocean). The atmosphere over the warmer surface has a decreased thermal stability, which causes a low level converging flow. Thus, the air parcels move horizontally toward the center of the heated region, where they rises upward. 
The characteristic flow pattern in Fig~\ref{fig:sb} is observed from our numerical simulation. The comparison in Fig~\ref{fig:sb} indicates that our numerical model simulates phenomena that is also observed in the nature. The streamline plot in Fig~\ref{fig:sb}$(b)$ shows that air moves to the left in the lower portion of the boundary layer above $x>0$, and the air in the upper portion moves to the right. This flow pattern is computed at $t=6.5$~h from the simulation with $H_0=231.48\hbox{ W m}^{-2}$. 


\subsection{Turbulent mixed layer formation}
\begin{figure*}[t]
  \centering
  \begin{tabular}{cc}
    $(a)~t=1$~h&    $(b)~t=2$~h\\
    \includegraphics[trim=0cm 3.8cm 0cm 3.5cm,clip=true,width=8cm]{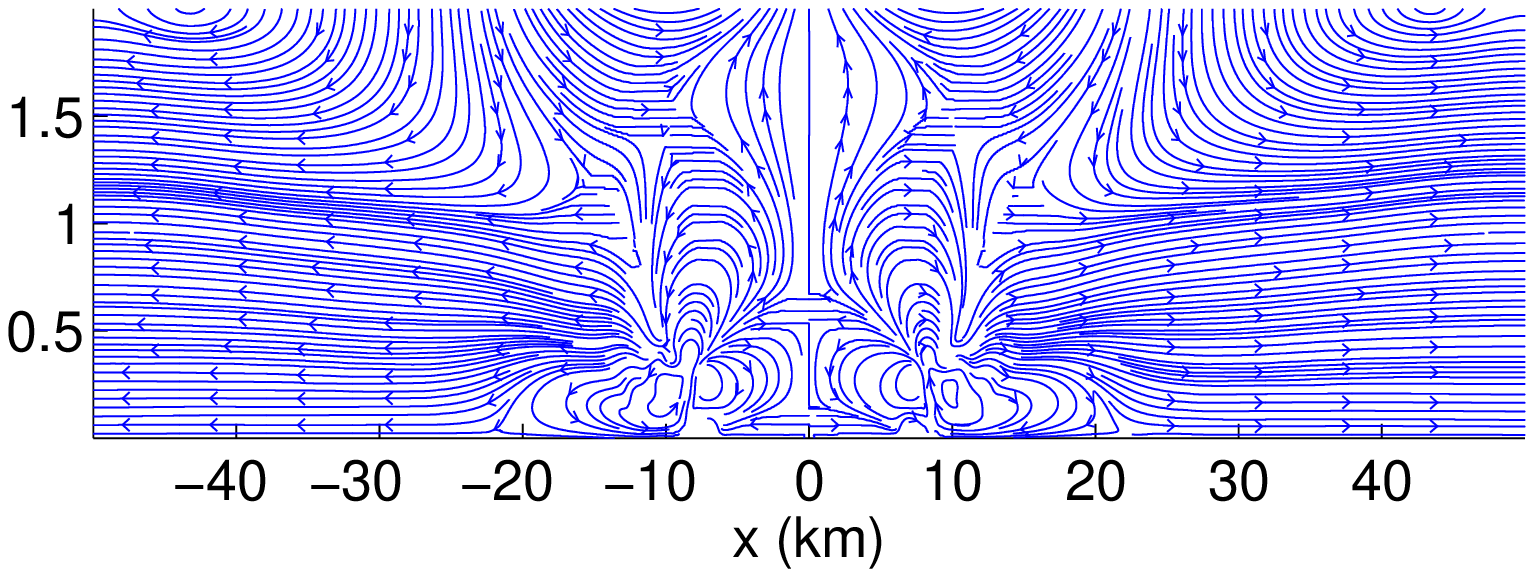}&
    \includegraphics[trim=0cm 3.8cm 0cm 3.5cm,clip=true,width=8cm]{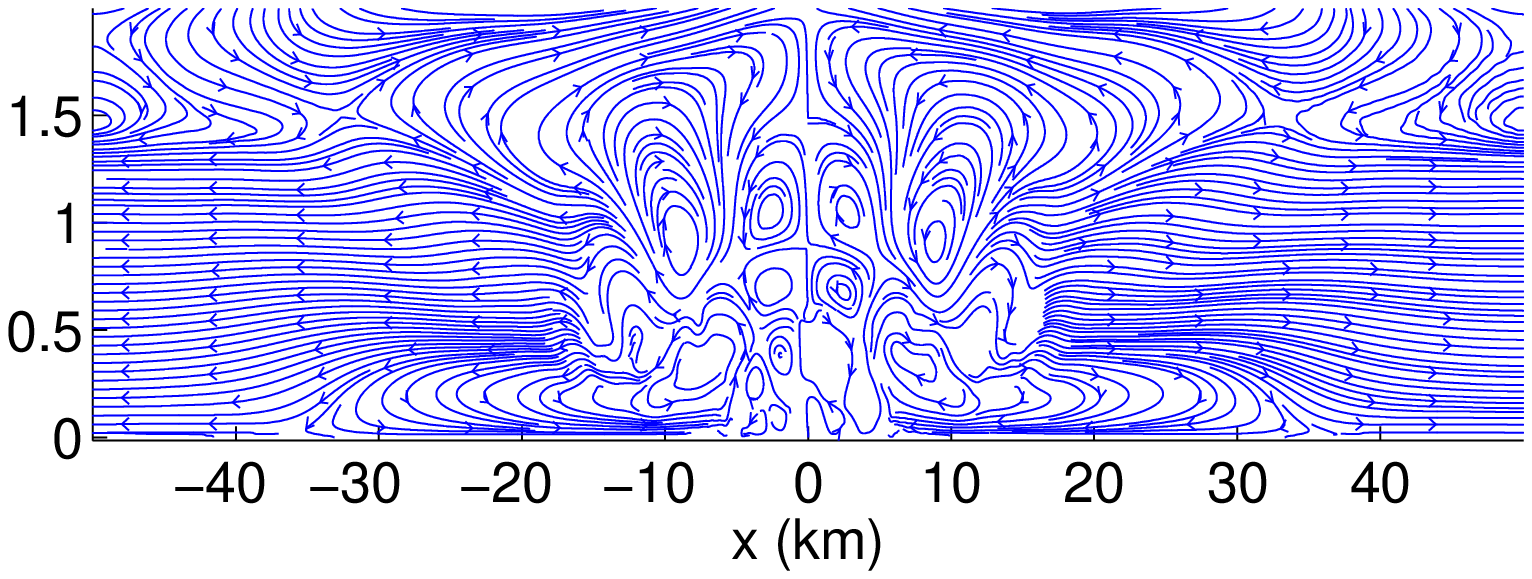}\\
    $(c)~t=4$~h&    $(d)~t=6$~h\\
    \includegraphics[trim=0cm 2.5cm 0cm 3.5cm,clip=true,width=8cm]{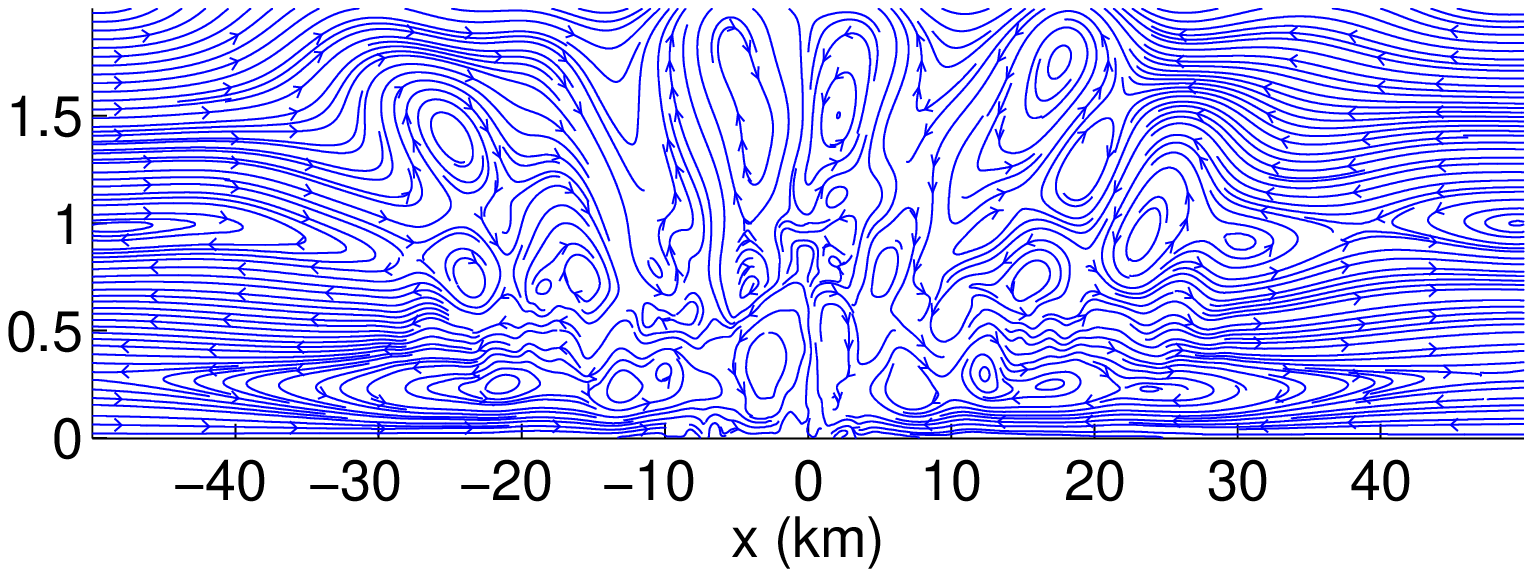}&
    \includegraphics[trim=0cm 2.5cm 0cm 3.5cm,clip=true,width=8cm]{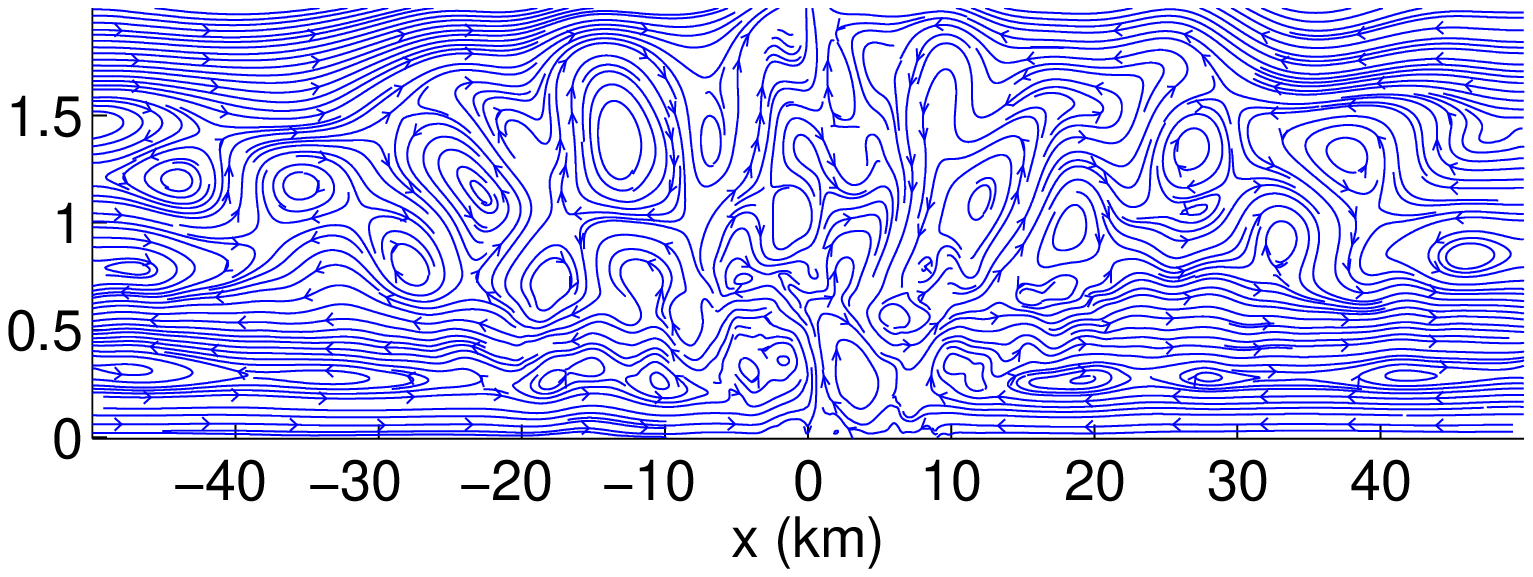}\\
  \end{tabular}
  \caption{The time evolution of span-wise vorticity is demonstrated using the streamline at $t=1$~h, $t=2$~h, $t=4$~h, and $t=6$~h for the simulation with $H_0=925.92\hbox{ W m}^{-2}$.}
  \label{fig:vor}
\end{figure*}

To understand the mixing and turbulent transport in the absence of shear or mechanically driven turbulence, we have initialized the flow with a stable stratification and no background wind everywhere except an unstable stratification is introduced near the surface. The surface Richardson number $Ri_0=(\frac{g}{\theta_0}\frac{\partial\theta}{\partial z}|_{z=0})H^2/U^2$ controls the initial strength of unstable stratification. The flow is initialized with $Ri_0=-1$, and allowed to evolve naturally with time. Thus, the resulting circulations characterize penetrative turbulent convection~\cite[][]{Deardorff69}. The time evolution of the streamlines in the entire domain is shown in Fig~\ref{fig:vor}. 
%

The relative surface heat flux for $-10\le x\le 10$ is $H_0=925.92\hbox{ W m}^{-2}$ ($\sim 0.75\hbox{ K m s}^{-1}$), where the background potential temperature $\bar\theta(z)$ in the stable region has a gradient $10^{\hbox{\tiny o}}/1$~km, and the buoyancy frequency is $N\approx 10^{-2}\hbox{s}^{-1}$. As seen from Fig~\ref{fig:vor}, when the sun heats the surface, eddies begin to form and rise upward. However, they return back downward due to stable stratification in the upper atmosphere. This generates an unstable mixing layer that is adjacent to the surface. This turbulent mixed layer underlies a stable region aloft. Based on the mean vertical profile of potential temperature (not shown), the depth of the mixed layer for this simulation is approximately $800$~m.

When an eddy loses buoyancy, it rises upward, and a negative horizontal buoyancy gradient occurs near its left edge, which generates a cyclonic circulation; similarly, anticyclonic circulation is formed on the other edge of the eddy~\cite[][]{Lane2008,Alam2011}. The pattern of such a circulation and the time evolution of the associated span-wise vorticity, during a penetrative turbulent convection, is realized from the streamlines in Fig~\ref{fig:vor}. This vortical pattern play its role as a heat transfer agent, thereby causing a imbalance between the buoyancy force and gravitational force. An important question of meteorological interest is whether the process leads to horizontal turbulent fluctuation, and if such fluctuation affects local weather prediction. At $t=1$~h, cyclonic/anticyclonic eddies have been formed near the outer edges of the heating region, and have reached a height of about $500$~m. Later, horizontal convection is observed; {\em i.e.} turbulent eddies move horizontally, where turbulent eddies reach a maximum vertical height of about $800$~m. Entrainment/detrainment occurs above this height.

\subsection{Temporal oscillation and downscale energy cascade}
\begin{figure*}
  \centering
  \begin{tabular}{cc}
    $H_0=462.96~\unit{W m}^{-2}$, $z=62.5~\unit{m}$&
    $H_0=925.92~\unit{W m}^{-2}$, $z=62.5~\unit{m}$\\
    \includegraphics[trim=0cm 2cm 0cm 0cm,clip=true,width=7cm,height=4.5cm]{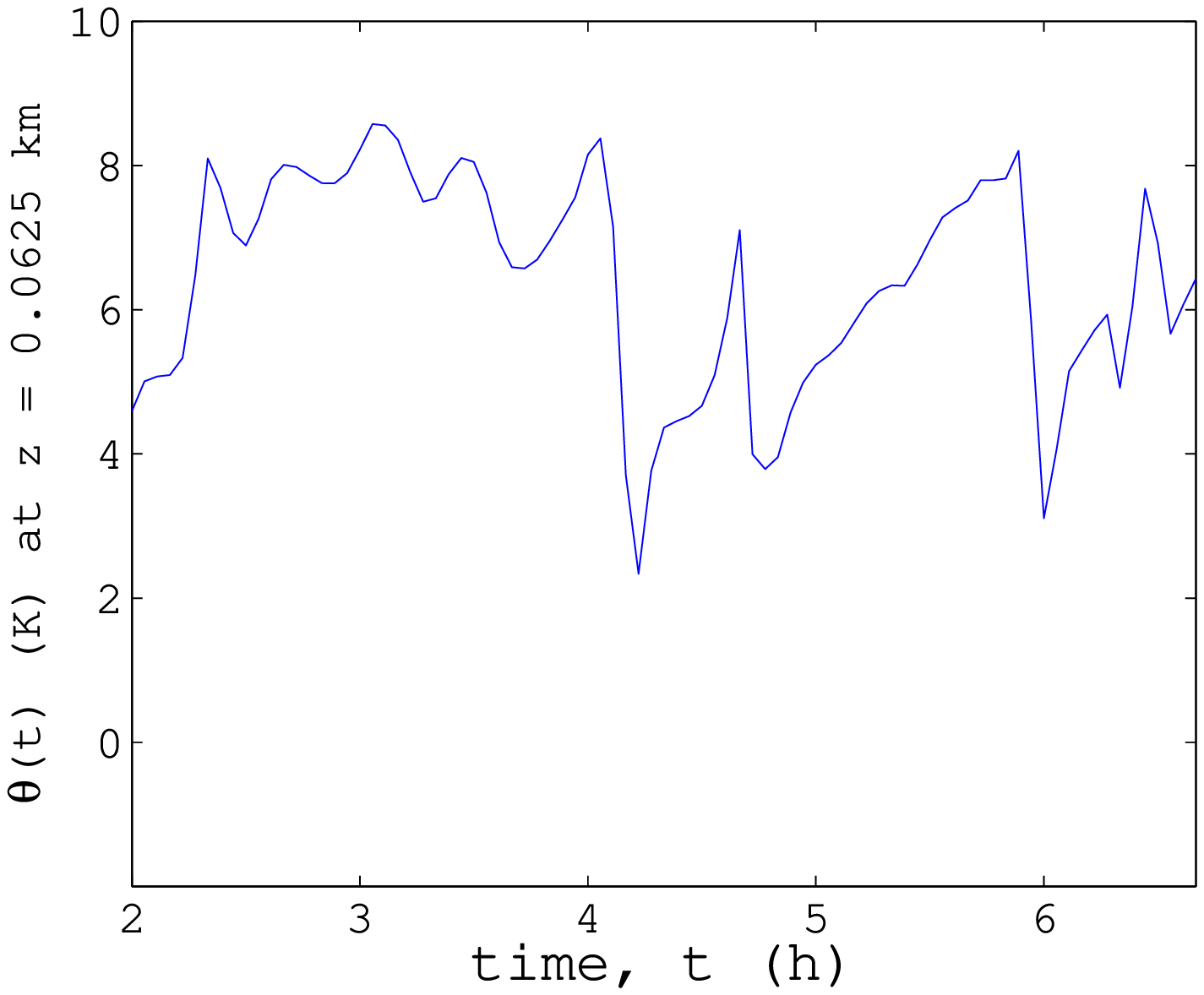}&
    \includegraphics[trim=0cm 0cm 0cm 0cm,clip=true,width=7cm,height=4.5cm]{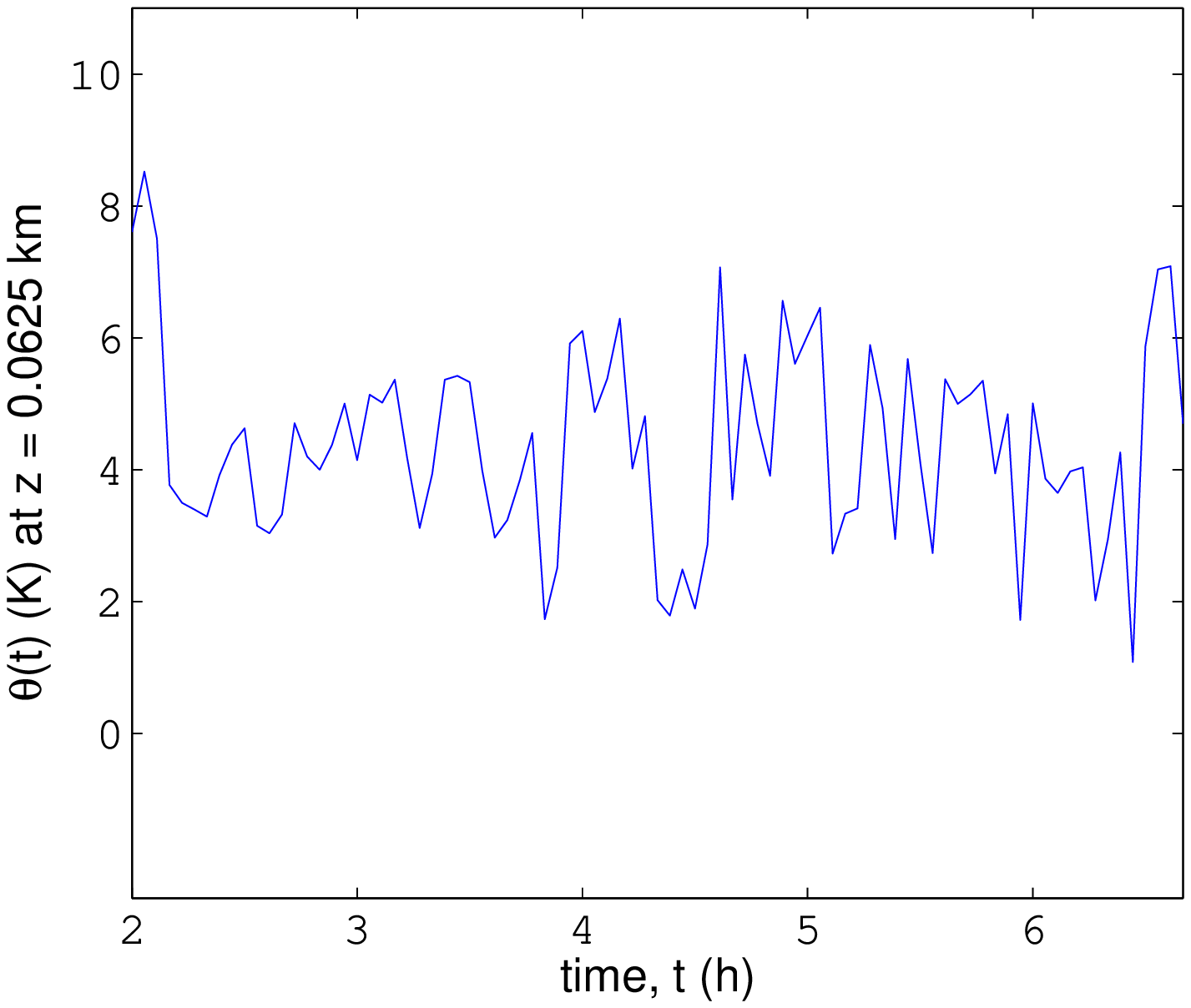}\\
    $H_0=462.96~\unit{W m}^{-2}$, $z=500~\unit{m}$&
    $H_0=925.92~\unit{W m}^{-2}$, $z=500~\unit{m}$\\
    \includegraphics[trim=0cm 0cm 0cm 0cm,clip=true,width=7cm,height=4.5cm]{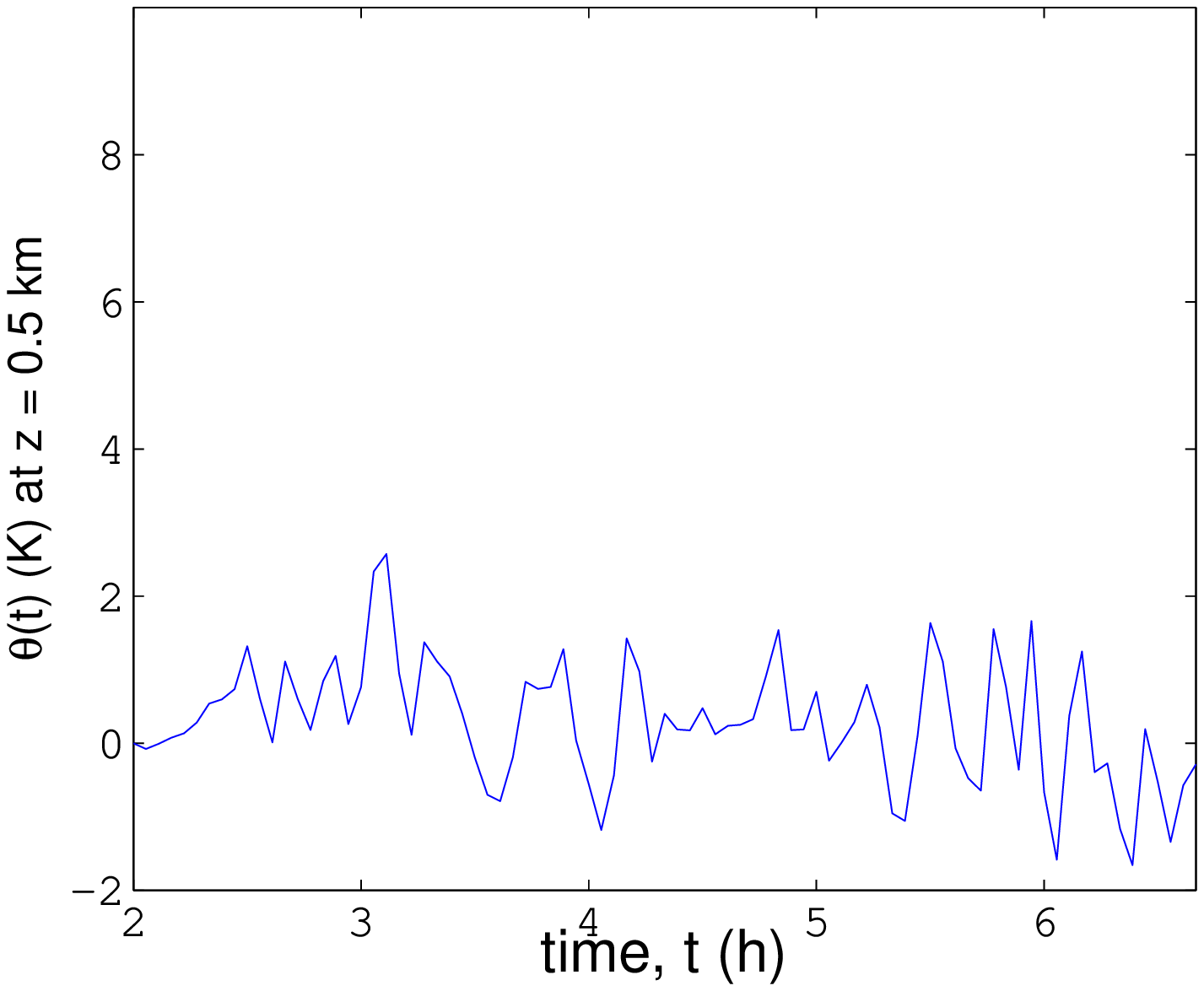}&
    \includegraphics[trim=0cm 0cm 0cm 0cm,clip=true,width=7cm,height=4.5cm]{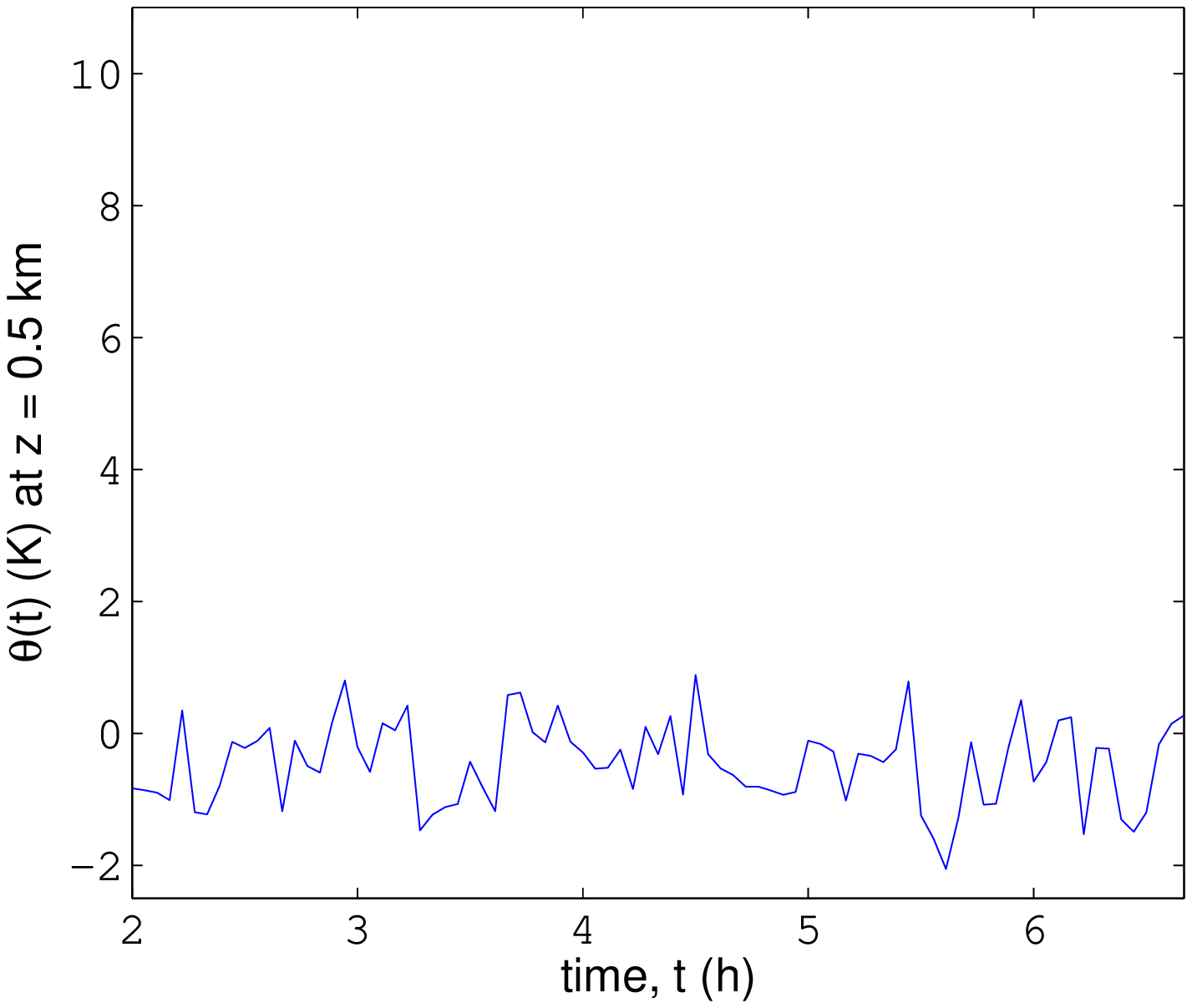}\\
  \end{tabular}
  \caption{Time series of potential temperature $\theta(0,z,t)$ near the surface ($z=62.5$~m) and near top of the mixing layer ($z=500$~m).}
  \label{fig:tst}
\end{figure*}

According to the Taylor's hypothesis, if turbulent statistics is approximately stationary and homogeneous, then the turbulent field is advected over the time scales of interest. Under this hypothesis, time series of potential temperature and horizontal velocity, as shown in Fig~\ref{fig:tst}, indicate that surface heterogeneity on a scale $\mathcal O(20~\unit{km})$ contributes toward downscale energy cascade. The growth of the horizontal potential temperature gradient develops a horizontal pressure gradient, which in turn generates horizontal wind. The time series of $\theta$ and $u$ at several vertical locations have been analyzed to characterize penetrative turbulence. 

\begin{figure*}
  \centering
  \begin{tabular}{cc}
    $H_0=462.96~\unit{W m}^{-2}$&
    $H_0=925.92~\unit{W m}^{-2}$\\ 
    \includegraphics[trim=0cm 0cm 2cm 0cm,clip=true,width=8cm]{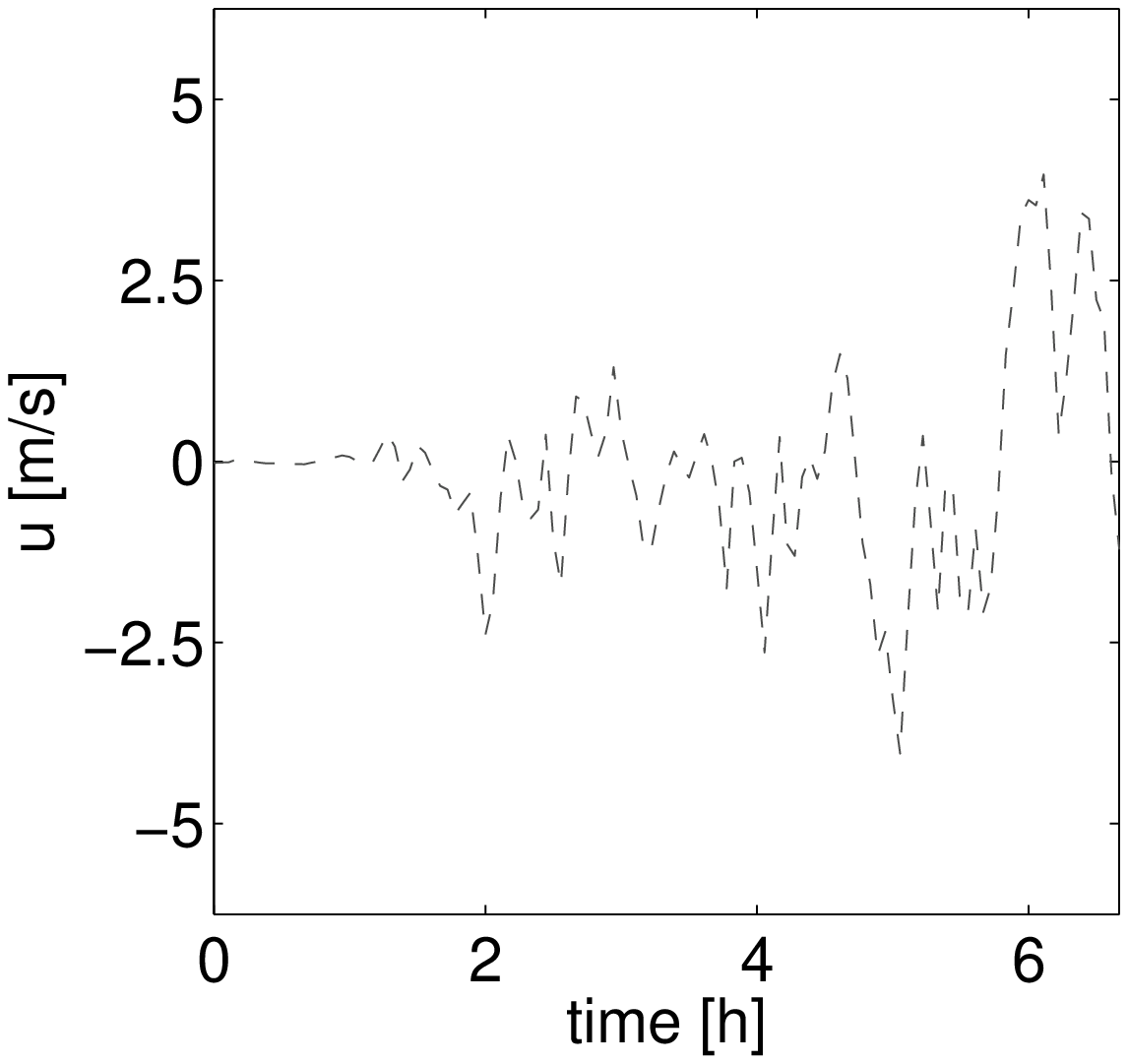}&
    \includegraphics[trim=0cm 0cm 2cm 0cm,clip=true,width=8cm]{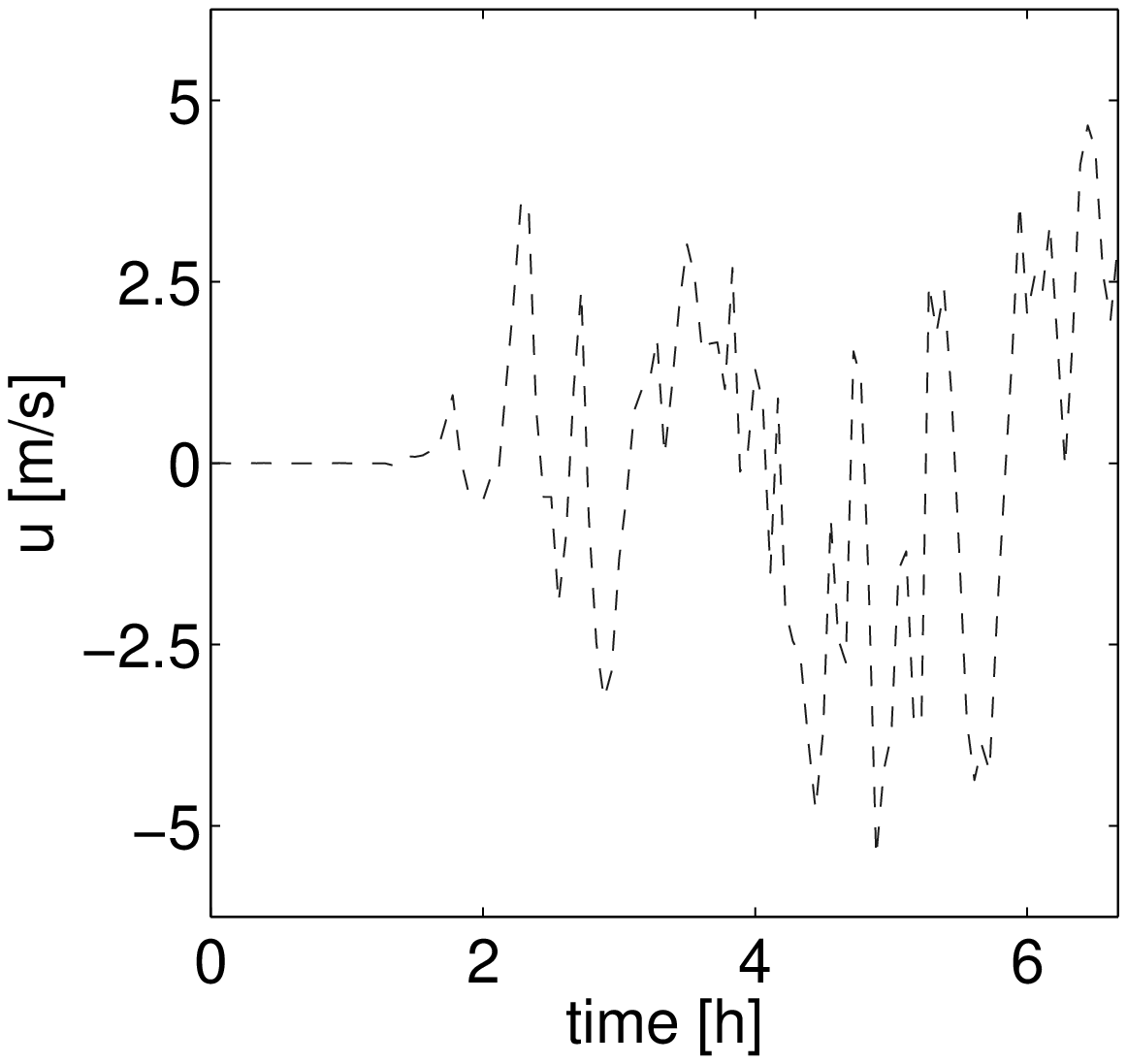}\\
  \end{tabular}
  \caption{The onset of turbulent fluctuation is shown through time series of horizontal velocity at $z=100~\unit{m}$ from the surface.}
  \label{fig:tsu}
\end{figure*}

The sensitivity of surface heat flux on the temporal fluctuation of $\theta$ is clear from Fig~\ref{fig:tst}. At $z=62.5$~m, frequency of oscillation increased at $H_0=925.92~\unit{W m}^{-2}$ compared to $H_0=462.96~\unit{W m}^{-2}$. However, at $z=500$~m, turbulence is seen fully developed in both cases. This indicates that turbulent kinetic energy cascades downscale as the energy is transported by internal waves. To verify that air flow over the simulated city is characterized by a horizontal flow of numerous rotating eddies, we present the horizontal velocity $u$, which is the wind component that is parallel to the direction of the surface heat flux variation. The onset of turbulent fluctuations is compared between two surfaces fluxes in Fig~\ref{fig:tsu}.  

\section{Concluding remarks}
In this article, we report some aspects of penetrative turbulence in an idealized daytime boundary layer over a city that is surrounded by rural areas. As internal waves excite from the interface between the mixed layer and stable layer, kinetic energy of boundary layer turbulence is transferred to upper label free atmosphere. A complete understanding of this mechanism remains a challenging future research topic. However, in the present article, we have shown that energy cascades downscale as turbulence penetrates upward. 
Our simulations demonstrate that characteristics of the induced horizontal flow may be significantly different depending on the surface heterogeneity from the perspective of the turbulence that is generated. 

The simulation is idealized. However, we have studied characteristics of horizontal flow with surface heat-flux variation on the scale of $20~\unit{km}$, which is the typical grid spacing in numerical weather prediction models. One of our objectives is to understand if thermally induced mesoscale perturbation has a connection to penetrative turbulence. In other words, we show that differential heating on the same label leads to a sustained turbulent flow. Note that we have compromised the three-dimensional simulation with a two-dimensional one in order to capture the energy containing large eddies. These results suggest that surface heterogeneity induced differential heating causes subgrid scale turbulent fluctuations, and the impact of surface heterogeneity could be more substantial on mesoscale atmospheric flows. In future studies, we plan to continue fully three-dimensional LES of penetrative turbulence. We plan to investigate how temporal oscillations characterize turbulent energy cascade, particularly with background wind conditions and rough surface.

\bibliographystyle{apalike}
\bibliography{bibrefs}
\end{document}